\documentclass[conference]{IEEEtran}
\IEEEoverridecommandlockouts
\usepackage{amsmath,amssymb,amsfonts}
\usepackage[linesnumbered,ruled,vlined]{algorithm2e}
\usepackage{algpseudocode}
\usepackage{graphicx}
\usepackage{textcomp}
\usepackage{xcolor}
\usepackage{url}
\begin{document}

\title{Interpretable Automatic Rosacea Detection with Whitened Cosine Similarity\\

}

\author{\IEEEauthorblockN{1\textsuperscript{st} Chengyu Yang}
\IEEEauthorblockA{\textit{Department of Computer Science} \\
\textit{New Jersey Institute of Technology}\\
Newark, New Jersey, USA \\
cy322@njit.edu}
\and
\IEEEauthorblockN{2\textsuperscript{nd} Chengjun Liu}
\IEEEauthorblockA{\textit{Department of Computer Science} \\
\textit{New Jersey Institute of Technology}\\
Newark, New Jersey, USA \\
chengjun.liu@njit.edu}
% \and
% \and
% \and
% \and
}

\maketitle

\begin{abstract}
According to the National Rosacea Society, approximately 16 million Americans suffer from rosacea, a common skin condition that causes flushing or long-term redness on a person’s face. To increase rosacea awareness and to better assist physicians to make diagnosis on this disease, we propose an interpretable automatic rosacea detection method based on whitened cosine similarity in this paper. The contributions of the proposed methods are three-fold. First, the proposed method can automatically distinguish patients suffering from rosacea from people who are clean of this disease with a significantly higher accuracy than other methods in unseen test data, including both classical deep learning and statistical methods. Second, the proposed method addresses the interpretability issue by measuring the similarity between the test sample and the means of two classes, namely the rosacea class versus the normal class, which allows both medical professionals and patients to understand and trust the results. And finally, the proposed methods will not only help increase awareness of rosacea in the general population, but will also help remind patients who suffer from this disease of possible early treatment, as rosacea is more treatable in its early stages. The code and data are available at \url{https://github.com/chengyuyang-njit/ICCRD-2025}.
\end{abstract}


\section{Introduction}
Rosacea is a commonly encountered chronic inflammatory skin disease in adults with a predilection for highly visible areas of the skin such as the face. It is characterized by flushing, redness, pimples, pustules and dilated blood vessels \cite{mikkelsen2016rosacea}. Rosacea symptoms often exhibit a cyclical pattern, with flare-ups lasting weeks to months before subsiding temporarily. In its early stages, however, rosacea may manifest as mild, transient facial redness (flushing), which can be easily overlooked. As a result, many patients may not medical attention until the condition has advanced, complicating treatment and management. Therefore, early detection is of extreme importance to facilitate timely intervention and improve clinical outcomes. Towards that end, we propose an interpretable automatic rosacea detection method based upon whitened cosine similarity.

The application of supervised deep learning methods to medical images, such as facial images of patients with rosacea, often encounters the challenge of insufficient labeled training data \cite{dhar2023challenges}. This issue is further complicated by the need to protect patient confidentiality, which limits data accessibility. Additionally, the rarity of certain diseases makes it even more difficult to gather an adequate amount of training data.

While research on rosacea and related skin conditions has been conducted using machine learning and computer vision/deep learning algorithms, most high-performing studies leveraging deep learning used datasets with nearly 10,000 images or even more \cite{thomsen2020deep}\cite{zhao2021novel}\cite{wu2020deep}\cite{zhu2021deep}. However, the datasets in these works are fully confidential, making them difficult to reproduce \cite{mohanty2022skin}.

As a result, deep learning methods tend to overfit the limited training data, which often leads to biased representation of the population distribution \cite{yang2024increasing}. To solve the problem of lacking enough training data for rosacea, research has been conducted to generate more rosacea patients' frontal facial images \cite{mohanty2022towards} leveraging the generative adversarial networks (GAN) \cite{goodfellow2020generative}. 

To simulate the real situation, that is, the amount of training data is limited and hard to collect, and also with the purpose of protecting patients' privacy, these generated rosacea patients' frontal facial images have been used to test the performance of some deep learning and statistical learning method \cite{yang2024increasing}. According to the study, the reported accuracy of both methods on the real unseen test data is only 89.5\% when the models are trained on these generated images. More specifically, the ResNet-18 method's recall rate is only 0.58. Which means 42\% patients who are actually suffering from rosacea could not be diagnosed due to deep learning method's problem of overfitting to the training data and lack of generalizability to the unseen data. Also, the dark-box nature of deep learning method makes it lack of interpretability and less reliable from the perspective of medical professionals. Even though the statistical method using PCA proposed in the study achieves a relatively higher recall rate, the precision rate is only 0.75. If implemented in real medical environments, this method could generate a substantial number of false positives, potentially overwhelming clinicians with unnecessary alerts. 

In this paper, we propose an interpretable automatic rosacea detection method leveraging the whitened cosine similarity to address the issues of both explainability and limitation of the training data. The contributions of our proposed method is three-fold. First, the proposed method effectively distinguishes patients with rosacea from healthy individuals with significantly better performance than existing approaches in terms of the accuracy, recall rate and precision rate, etc. when tested on the unseen data. Second, it tackles the issue of interpretability by assessing the similarity between the test sample and the centroids of two distinct classes, rosacea and normal, allowing both medical practitioners and patients to easily understand and trust the outcomes. Finally, this method not only raises awareness of rosacea among the general population but also encourages timely early treatment, as rosacea is more manageable when diagnosed at earlier stages. 

\section{Method}
We now present our automatic rosacea detection based upon whitened cosine similarity. It utilizes the means of the two classes, namely, the rosacea vs. the normal class, and measures the similarity between each test sample and the means of the two classes, giving an interpretable perspective for both the patients and the medical professionals.

Similarity measures play an important role in pattern recognition and computer vision \cite{bowyer1998empirical}. Whitened Cosine Similarity is a refined variation of cosine similarity that addresses certain limitations in traditional cosine similarity by incorporating data whitening techniques. It is particularly useful in applications involving high-dimensional data, where redundant or correlated features can impact the performance of similarity measures. It has been robustly demonstrated for superior pattern recognition performance on large scale experiments \cite{moon2001computational}. It has also been shown that the whitened cosine similarity measure is related to the Bayes decision rule for minimum error under some specific assumptions \cite{liu2007bayes}. 

Let \( \delta_{WC} \) represent the whitened cosine
similarity measure, which may be formulated as follows:

\begin{equation}\label{eq1}
\delta_{WC}(\mathcal{U}, \mathcal{V}) = \frac{({W}^{t}\mathcal{U})^{t}({W}^{t}\mathcal{V})}{\|{W}^{t}\mathcal{U}\|\|{W}^{t}\mathcal{V}\|}
\end{equation}
where $\mathcal{U,V} \in \mathbb{R}^{d}$ are two pattern vectors, $\|\cdot\|$ denotes the norm operator, and ${W}$ is the whitening transformation matrix, which may be specified by means of the covariance matrix. The covariance matrix of all samples, \( \Sigma = \mathcal{E}\{(\mathcal{X} - M_0)(\mathcal{X} - M_0)^{t}\} \), can be factorized using principal component analysis (PCA): \( \Sigma = \Phi \Lambda \Phi^{t} \), where \( \mathcal{E}(\cdot) \) is the expectation operator, \(M_0\)  is the grand mean vector, \(\Phi\) is an orthogonal eigenvector matrix, and \(\Lambda\) a diagonal eigenvalue matrix. The transformation, \( W = \Phi \Lambda^{-1/2} \) is called the whitening transformation \cite{fukunaga2013introduction}. Note that, under this particular whitening transformation, (\ref{eq1}) becomes

\begin{equation}\label{eq2}
\delta_{WC}(\mathcal{U}, \mathcal{V}) = \frac{\mathcal{U}^{t}\Sigma^{-1}\mathcal{V}}{\|{W}^{t}\mathcal{U}\|\|{W}^{t}\mathcal{V}\|}
\end{equation}
which is the commonly used form of the whitened cosine similarity measure, but it is not widely used in practice due to the complication of computing the inverse of the covariance matrix.

Our method leverages this similarity measure to build an intuitive and interpretable automatic rosacea detector. For each test sample, we then calculate the whitened cosine similarity between this test sample and the means of the two classes: normal faces and faces having rosacea. If the similarity between the test sample and the mean of normal faces is larger than that between the test sample and the mean of faces having rosacea, the test sample is classified to be normal, otherwise, it is classified to be rosacea. The detailed algorithm is presented in Algorithm \ref{alg}.

\begin{algorithm}\label{alg}
    \caption{Automatic Rosacea Detection with Whitened Cosine Similarity}

    \DontPrintSemicolon
    \SetAlgoLined
    \SetNoFillComment
    \LinesNotNumbered

    \KwIn{$\mathbf{X_{\textit{normal}}}\in \mathbb{R}^{d\times n}, \mathbf{X_{\textit{rosacea}}}\in \mathbb{R}^{d\times m}, \mathcal{X} \in \mathbb{R}^{d}$ }
    \KwOut{\textit{True / False}}
    \tcc{\textsc{get one matrix of all samples}}
    $\mathbf{X} \in \mathbb{R}^{d
    \times (n+m)}\gets [\mathbf{X_{\textit{normal}}}, \mathbf{X_{\textit{rosacea}}}]$\;
    \;
    
    \tcc{\textsc{class mean}}
    $M_{\textit{normal}} \gets {(\sum_{i=1}^{n}\mathbf{X}[\ :\ ,i] )}\ /\ n  $\;
    $M_{\textit{rosacea}} \gets {(\sum_{i=n+1}^{n+m}\mathbf{X}[\ :\ ,i] )}\ /\ m  $\;
    \;
    
    \tcc{\textsc{all samples' mean}}
    $M_0 \gets {(\sum_{i=1}^{n+m}\mathbf{X}[\ :\ ,i] )}\ /\ (n+m)  $\;
    \;

    \tcc{\textsc{center the training data}}
    \For{$i=1$ \KwTo $(n+m)$}{
    $\mathbf{X}[\ :\ ,i] \gets \mathbf{X}[\ :\ ,i] - M_0$
    }
    \;

    \tcc{\textsc{unbiased estimate of covariance}}
    $\Sigma \gets \mathbf{X}\mathbf{X}^{t}/ (n+m-1)$ \;
    \;

    \tcc{\textsc{PCA on covariance matrix}}
    $\Sigma = \Phi \Lambda \Phi^{t}$\;
    \;

    \tcc{\textsc{whitening transformation matrix}}
    $W \gets \Phi \Lambda^{-1/2}$\;
    \;

    \tcc{\textsc{compare and classify}}
    \eIf{$\delta_{WC}(\mathcal{X}, M_{\text{normal}}) < \delta_{WC}(\mathcal{X}, M_{\text{rosacea}})$}{
    \text{return} True
    }{
    \text{return} False
    }

\end{algorithm}

We then compared our method with the deep learning and statistical learning method used in \cite{yang2024increasing}
on the same dataset. Other than that, we also implemented dimension reduction and feature extraction techniques like principal component analysis (PCA), and KNN \cite{fix1985discriminatory} with different distance metrics (L1, L2, cosine similarity) for the purpose of comparison. K-Nearest Neighbors (KNN) is a simple, intuitive, and non-parametric machine learning algorithm used for both classification and regression tasks. It relies on the principle of similarity between data points and does not involve explicit training of a model. The similarity between data points can be measured by Manhattan distance (L1), which is given by 

\begin{equation}
d_{\text{Manhattan}}(\mathbf{x}, \mathbf{y}) = \sum_{i=1}^{n} |x_i - y_i|
\end{equation}
Or Euclidean distance (L2), which is given by

\begin{equation}
d_{\text{Euclidean}}(\mathbf{x}, \mathbf{y}) = \sqrt{\sum_{i=1}^{n} (x_i - y_i)^2}
\end{equation}
Or cosine similarity, which is given by
\begin{equation}
\text{Cosine Similarity} = \frac{\mathbf{x} \cdot \mathbf{y}}{\|\mathbf{x}\| \|\mathbf{y}\|}
\end{equation}

\section{Experiment}
In this section, we first introduce the dataset, then we disclose some technical details which are especially important and helpful when implementing the algorithm \ref{alg} under the circumstances when the dimension of the data is extremely high. In the end, we compare and analyze the performance of our method with that of other classic deep learning and statistical learning methods in terms of metrics like accuracy, precision rate and recall rate, etc.

\subsection{Dataset}
We use the same dataset as is used in \cite{yang2024increasing}, for the same purpose of simulating the real situation where the amount of training data is limited and hard to collect, and also to protect patients' privacy.

\begin{figure*}
\includegraphics[width=\textwidth]{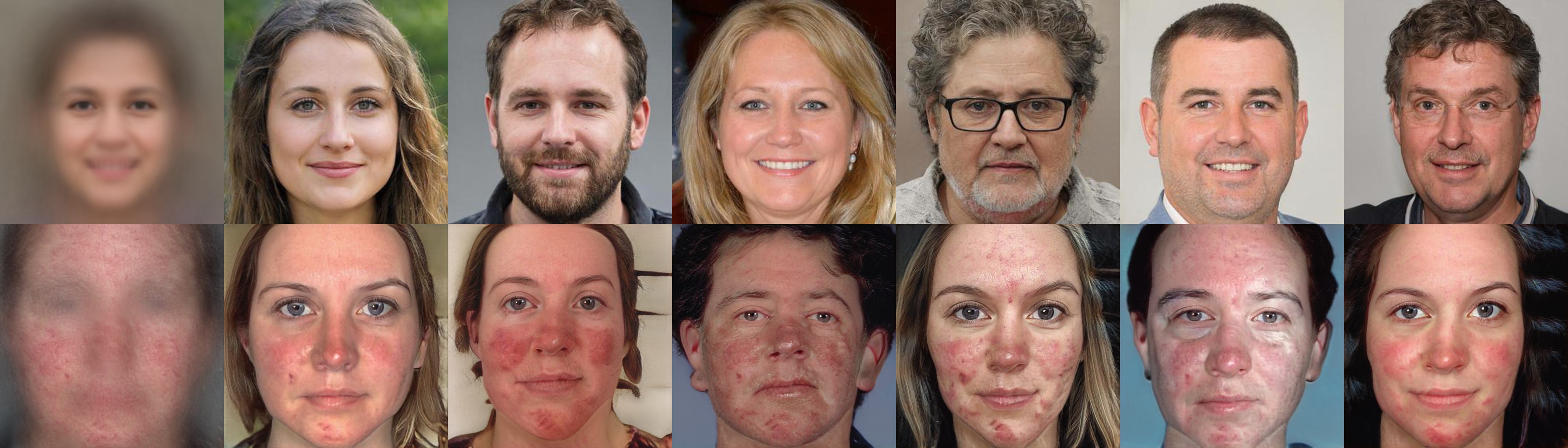}
\caption{The first column shows the mean images of the two training datasets from the rosacea negative and positive classes, respectively. The remaining six columns display six pair of random example images from the two training datasets corresponding to the normal people and those with rosacea, respectively.} \label{fig1}
\end{figure*}

\subsubsection{Training and Validation Dataset}
We use generated images for both rosacea positive and negative cases. As is shown in Fig \ref{fig1}, the frontal face images with rosacea have been generated using the GAN \cite{mohanty2022towards}\cite{goodfellow2020generative}. We therefore use 300 of the generated rosacea frontal face images and split them into the training set with 250 images and the validation set with the remaining 50 images. For the rosacea negative cases, we use 600 frontal face images generated from the Style-GAN \cite{karras2019style} and split them into a training set with 500 images and the validation set with the remaining 100 images.
\subsubsection{Test Dataset} To assess the performance of the proposed automatic rosacea detection method on real images, we use 50 real frontal face images with
rosacea from various sources including Kaggle, DermNet, and National Rosacea Society. These images are selected with a standard of having at least 200 x 200 pixel resolution. The face in each image is detected and aligned using the DeepFace \cite{taigman2014deepface} and cropped and resized to the same size with the training and the validation images which are $512 \times 512 \times 3 $ for the purpose of testing. In addition, we include 150 real people's frontal faces not having rasacea from the CelebA Dataset\cite{liu2015deep} in our test data set. 

\subsection{Technical Details Disclosure}
When implementing the algorithm, the dimension of the data determines the size of the empirical covariance matrix. If the dimension of the data is very high ($d \gg n + m$), for example $512 \times 512 \times 3$ in our case, then it will be extremely time consuming and memory inefficient to perform PCA on the covariance matrix. 

Instead of performing PCA on $\Sigma$, which is a matrix of dimension $d \times d$ given by $\mathbf{X}\mathbf{X}^{t}$ multiplied by a constant $\frac{1}{n+m-1}$ that we ignore for now, we perform PCA on $\mathbf{X}^{t}\mathbf{X}$, which only has a dimension of ${(n + m)\times(n+m)}$. Let $v$ be an eigenvector of $\mathbf{X}^{t}\mathbf{X}$ with eigenvalue $\lambda$, then

\begin{equation}\label{eq3}
    \mathbf{X}^{t}\mathbf{X}v = \lambda v
\end{equation}
Left multiply both sides of (\ref{eq3}) with matrix $\mathbf{X}$, we get

\begin{equation}
    \mathbf{X} (\mathbf{X}^{t}\mathbf{X}v )= \mathbf{X} (\lambda v)
\end{equation}
After applying matrix product associativity, we get

\begin{equation}
 (\mathbf{X} \mathbf{X}^{t})\mathbf{X}v = \lambda (\mathbf{X} v)
\end{equation}

In other words, if $v$ is an eigenvector of $\mathbf{X}^{t}\mathbf{X}$, then $\mathbf{X}v$ is an eigenvector of $\mathbf{X}\mathbf{X}^{t}$ with the same eigenvalue. Therefore, we obtain the PCA on the much larger original empirical covariance matrix. It should also be noted that if $v$ is a normalized eigenvector of $\mathbf{X}^{t}\mathbf{X}$, $\mathbf{X}v$ is not necessarily a normalized eigenvector of $\mathbf{X}\mathbf{X}^{t}$.

\subsection{Performance Evaluation and Comparison}
\subsubsection{Evaluation Metrics}
In this section, we introduce the evaluation metrics that we are going to use. To be more specific, we use accuracy, recall rate, precision rate, and F1 score. For a single method, all these four metrics are the higher, the better. The definitions of these four metrics are given below, where TP stands for the number of true positive predictions, TN stands for the number of true negative predictions, FP stands for the number of false positive predictions and FN stands for the number of false negative predictions. 
\begin{equation}
\text{Accuracy} = \frac{\text{TP} + \text{TN}}{\text{TP} + \text{TN} + \text{FP} + \text{FN}}
\end{equation}

\begin{equation}
\text{Recall} = \frac{\text{TP}}{\text{TP} + \text{FN}}
\end{equation}

\begin{equation}
\text{Precision} = \frac{\text{TP}}{\text{TP} + \text{FP}}
\end{equation}

\begin{equation}
\text{F1-Score} = 2 \cdot \frac{\text{Precision} \cdot \text{Recall}}{\text{Precision} + \text{Recall}}
\end{equation}

\begin{table}[t]
\caption{Different methods performance on validation data}
\centering
\begin{tabular}{|l|c|c|c|c|}
\hline
\textbf{Method} & \textbf{Accuracy} & \textbf{Recall} & \textbf{Precision}&  \textbf{F1}\\
\hline
ResNet-18 \cite{yang2024increasing}\cite{he2016deep} & 1.00& 1.00 & 1.00 & 1.00 \\
\hline
Class dependent PCA \cite{yang2024increasing}& 0.90 & 0.96 & 0.77 & 0.86 \\
\hline
Class independent PCA \cite{fukunaga2013introduction}& 0.97 & 0.98 & 0.94 & 0.96 \\
\hline
KNN with L1 metric \cite{fix1985discriminatory}& 0.99 & 1.00 & 0.98 & 0.99 \\
\hline
KNN with L2 metric \cite{fix1985discriminatory}& 0.99 & 1.00 & 0.98 & 0.99 \\
\hline
KNN with cosine metric \cite{fix1985discriminatory}& 1.00 & 1.00 & 1.00 & 1.00 \\
\hline
KNN-L2 after PCA \cite{fix1985discriminatory}& 0.99 & 1.0 & 0.98 & 0.99 \\
\hline
\hline
\textbf{Our Method} & \textbf{1.00} & \textbf{1.00} & \textbf{1.00} & \textbf{1.00} \\
\hline
\end{tabular}

\label{table1}
\end{table}

\begin{table}[t]
\caption{Different methods performance on test data}
\centering
\begin{tabular}{|l|c|c|c|c|}
\hline
\textbf{Method} & \textbf{Accuracy} & \textbf{Recall} & \textbf{Precision}&  \textbf{F1}\\
\hline
ResNet-18 \cite{yang2024increasing}\cite{he2016deep} & 0.90 & 0.58 & 1.00 & 0.73 \\
\hline
Class dependent PCA \cite{yang2024increasing}& 0.90 & 0.88 & 0.75 & 0.81 \\
\hline
Class independent PCA \cite{fukunaga2013introduction}& 0.93 & 0.84 & 0.875 & 0.86 \\
\hline
KNN with L1 metric \cite{fix1985discriminatory}& 0.96 & 0.88 & 0.96 & 0.92 \\
\hline
KNN with L2 metric \cite{fix1985discriminatory}& 0.95 & 0.88 & 0.90 & 0.89 \\
\hline
KNN with cosine metric \cite{fix1985discriminatory}&0.95 & 0.80 & 0.98 & 0.88 \\
\hline
KNN-L2 after PCA \cite{fix1985discriminatory}& 0.95 & 0.82 & 0.95 & 0.88 \\
\hline
\hline
\textbf{Our Method} & \textbf{0.99} & \textbf{0.96} & \textbf{1.00} & \textbf{0.98} \\
\hline
\end{tabular}

\label{table2}
\end{table}

\subsubsection{Performance Analysis}
As we can see from TABLE \ref{table1} and TABLE \ref{table2}, our method achieved 100\% accuracy on the validation data and 99\% accuracy on the test data, significantly outperformed other methods in terms of accuracy, recall rate, precision rate and the F1 score. In the confusion matrix of our method on test data shown in TABLE \ref{table3}, only two out of fifty patients who actually suffer from rosacea are not diagnosed by our method, leading to the highest recall rate of all the listed methods. A higher recall rate will not only help increase rosacea awareness in the general population but also help remind the patients who suffer from this disease of possible early treatment since rosacea is more treatable at its early stages.

It is also worth noting that, even though the classic deep learning method ResNet-18 achieves 100\% accuracy on the validation set, due to the limitation on the size and the generated nature of the training and validation samples, it suffers from the problem of overfitting, thus cannot generalize well to the real world test dataset. Actually, TABLE \ref{table2} shows that its recall rate is only 0.58, which means 21 out of 50 rosacea positive cases are reported to be negative. In this case, a lot of patients actually having rosacea will not be diagnosed and will miss the best opportunity to start early treatment. Furthermore, deep neural networks' dark-box nature makes the result more unreliable from the perspective of medical practitioners. 

\begin{table}[t]
\caption{Confusion matrix of our method on test data}
\centering
\begin{tabular}{|l|c|c|}
\hline
 & \textbf{Classified as Normal} & \textbf{Classified as Rosacea} \\
\hline
Actual Normal & 150 & 0 \\
\hline
Actual Rosacea & 2 & 48   \\
\hline
\end{tabular}

\label{table3}
\end{table}

\subsubsection{Insights}
Here we display the two rosacea test cases that our method misclassified as normal people's faces in Fig 2.

\begin{figure}
\centering
\includegraphics[width=150pt]{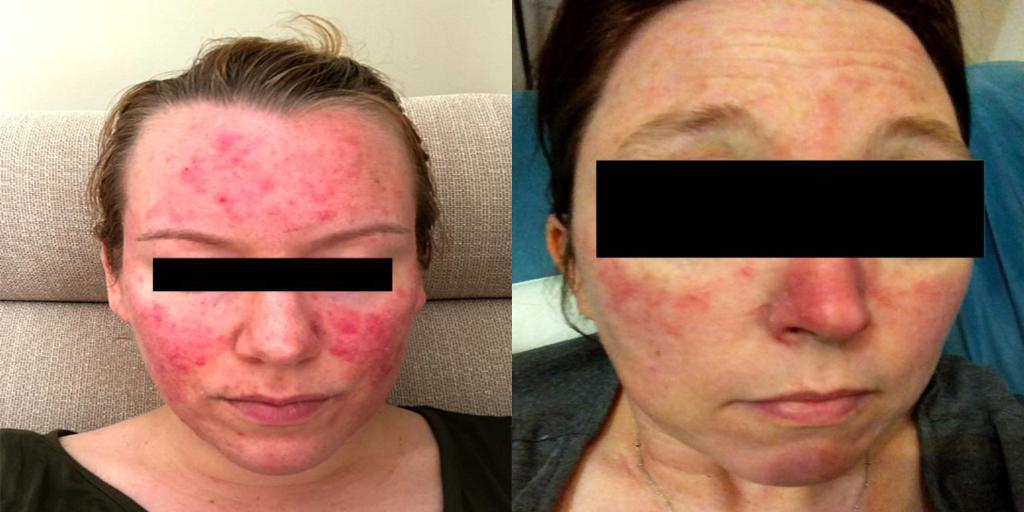}
\caption{The two test cases that our method misclassified.} \label{fig2}
\end{figure}

As we can see, for the first case, the face area is too small compared to the face sizes in the training dataset. For the second case, the black area covering the patient's eyes is too large, which may extort the main features of the test image.

\section{Conclusion}
In this paper, we propose a method for automatic rosacea detection utilizing whitened cosine similarity.  demonstrates significant advancements in both accuracy and interpretability compared to existing approaches. By effectively distinguishing between rosacea and normal cases, our method achieves a high recall rate, ensuring that the majority of patients suffering from rosacea are accurately diagnosed. This is particularly crucial given the treatability of rosacea in its early stages, which can lead to improved patient outcomes. Furthermore, the interpretability of our approach allows both medical practitioners and patients to understand and trust the diagnostic results, fostering greater awareness of rosacea and encouraging timely intervention. As the challenges of limited training data and the dark-box nature of deep learning methods persist, our research highlights the importance of developing transparent and reliable diagnostic tools in the field of medical imaging. Future work will focus on further refining this method and exploring its application in diverse clinical settings to enhance the management of rosacea and similar skin conditions.

\bibliographystyle{IEEEtran}
\bibliography{refs}

\end{document}